# Drone Carry-on Weight and Wind Flow Assessment via Micro-Doppler Analysis

Dmytro Vovchuk, Oleg Torgovitsky, Mykola Khobzei, Vladyslav Tkach, Sergey Geyman, Anton Kharchevskii, Andrey Sheleg, Toms Salgals, Vjaceslavs Bobrovs, Shai Gizach, Aviel Glam, Niv Haim Mizrahi, Alexander Liberzon, and Pavel Ginzburg

*Abstract*— Remote monitoring of drones has become a global objective due to emerging applications in national security and managing aerial delivery traffic. Despite their relatively small size, drones can carry significant payloads, which require monitoring, especially in cases of unauthorized transportation of dangerous goods. A drone's flight dynamics heavily depend on outdoor wind conditions and the carry-on weight, which affect the tilt angle of a drone's body and the rotation velocity of the blades. A surveillance radar can capture both effects, provided a sufficient signal-to-noise ratio for the received echoes and an adjusted postprocessing detection algorithm. Here, we conduct a systematic study to demonstrate that micro-Doppler analysis enables the disentanglement of the impacts of wind and weight on a hovering drone. The physics behind the effect is related to the flight controller, as the way the drone counteracts weight and wind differs. When the payload is balanced, it imposes an additional load symmetrically on all four rotors, causing them to rotate faster, thereby generating a blade-related micro-Doppler shift at a higher frequency. However, the impact of the wind is different. The wind attempts to displace the drone, and to counteract this, the drone tilts to the side. As a result, the forward and rear rotors rotate at different velocities to maintain the tilt angle of the drone body relative to the airflow direction. This causes the splitting in the micro-Doppler spectra. By performing a set of experiments in a controlled environment, specifically, an anechoic chamber for electromagnetic isolation and a wind tunnel for imposing deterministic wind conditions, we demonstrate that both wind and payload details can be extracted using a simple deterministic algorithm based on branching in the micro-Doppler spectra. In the context of machine learning algorithms used in radar science today, a deterministic, physics-based model offers significant value, not only by providing an additional layer of control but also due to its interpretability and physical grounding, which complements data-driven approaches.

## I. INTRODUCTION

D RONES become a resource for a wide range of applications over the past few decades, including but not limited to infrastructure monitoring, airborne remote sensing, logistics and delivery, rescue operations, and many others [1], [2], [3], [4], [5], [6], [7], [8], [9], [10], [11], [12], [13], [14]. Advanced navigation systems, high level of automatization, low cost, extended autonomous operation, and many other advantages promote extensive use of drones very soon and their impact will grow dramatically from year to year [15], [16], [17], [18], [19], [20], [21], [22]. However, by now, the majority of flights are still not subject to precise control, with operational responsibilities shifted to users, posing ongoing challenges in enforcing regulations, particularly in ensuring payload compliance and detecting unauthorized transport. Due to their low cost and unlicensed accessibility, small drones can be used by unauthorized users to carry dangerous items, spot on classified sites, interfere with air traffic, and for other undesired purposes.

Among the various monitoring systems available, the primary methods include camera, acoustic, and radar detection techniques, each with its own advantages and limitations [23], [24], [25], [26]. While high-resolution imaging can allow for highly accurate target recognition, it heavily relies on a line of sight, ambient illumination, and intensive signal processing, to name the key constraints. Acoustic approaches are rather range-limited and susceptible to environmental noises [27], [28]. Radar systems have already proven themselves to operate in harsh conditions, providing real-time, reliable detection of airborne targets [29], [30], [31], [32], [33].This is the reason why those systems keep developing and are actively deployed on many different platforms, including automotive.

In the radar realm, small drones are a relatively new class of airborne targets [34], [35]. Those targets have enormously

This work was supported by Department of the Navy, Office of Naval Research Global, under ONRG Award N62909–21–1–2038, Israel Science Foundation (ISF grant number 1115/23), Science Forefront (Israel), project 0006764. Israel Innovation Authority, NATO SPS project No. G6118. Niv Haim Mizrahi acknowledges the Israel Smart Transportation Centre and The Shlomo Shmelzer Institute for Smart Transportation. The RTU team acknowledges the support from the RRF project Latvian Quantum Technologies Initiative Nr. 2.3.1.1.i.0/1/22/I/CFLA/001 and the 1.1.1.9 Activity "Post-doctoral Research" Research application No 1.1.1.9/LZP/1/24/166 "Linear Industrial Monitoring System based on Hyperspectral Cameras and AI Algorithms (LIF-HYCAI)". *(Corresponding author: Dmytro Vovchuk, e-mail: dimavovchuk@gmail.com).* Dmytro Vovchuk, Oleg Torgovitsky, Mykola Khobzei, and Vladyslav Tkach are contributed equally to this work.

Dmytro Vovchuk, Mykola Khobzei, Vladyslav Tkach, Toms Salgals, and Vjaceslavs Bobrovs are with the Institute of Photonics, Electronics and Telecommunications, Riga Technical University, Azenes Street 12, Riga 1048, Latvia (e-mail: mykola.khobzei@rtu.lv, vladyslav.tkach@rtu.lv).

Dmytro Vovchuk, Oleg Torgovitsky, Sergey Geyman, Anton Kharchevskii, Andrey Sheleg, Shai Gizach, Aviel Glam, and Pavel Ginzburg are with the School of Electrical Engineering, Tel Aviv University, Tel Aviv 69978, Israel (e-mail: dimavovchuk@gmail.com, pginzburg@tauex.tau.ac.il).

Niv Haim Mizrahi, and Alexander Liberzon are with the School of Mechanical Engineering, Tel Aviv University, Tel Aviv 6779801, Israel.

Pavel Ginzburg is with the Center for Light–Matter Interaction, Tel-Aviv University, Tel-Aviv 6779801, Israel.

small radar cross-sections, hardly distinguishable from clutter [36]. For example, birds generate comparable radar signatures [37]. In this case, an additional target classification has to be performed. Micro-Doppler analysis, differentiating signatures of flapping wings and rotating blades comes at a rescue, though demanding a more accurate target investigation (e.g., time on target and signal-to-noise ratio (SNR) in detection) [33], [34], [38], [39], [40], [41], [42]. Furthermore, another critical aspect is a low flight latitude, which puts clutter-filtering aspects at the core, as ground reflection and multi-path interference become key factors.

Apart from detecting and identifying a drone, an important objective is to verify whether the item carries a payload and its weight [43]. This information can also be extracted from micro-Doppler signatures, as it was demonstrated in [30], [44], [45], [46], [47], [48], [49], [50]. The link between the payload and spectral signatures comes from the aerodynamics of the thrust and lift generated by the blades. In other words, the blades must rotate faster to lift a larger weight. However, several other phenomena can contribute to the same effect. Outdoor conditions, primarily involving wind, cause the drone to adjust its tilt angle to the wind direction and rotate its blades faster to keep hovering at a fixed position despite the wind. This complexity was acknowledged in a series of pioneering works that applied machine-learning techniques to extract payload information from micro-Doppler signatures [43], [51], [52], [53], [54]. Owing to the previously mentioned problem complexity, a reliable payload estimation algorithm was not found.

Here, we propose to disentangle the various effects contributing to the micro-Doppler spectrum and study them independently. This separation enables a higher degree of determinism, which can improve the performance of future classification algorithms. As a first step, we investigate a drone with a controllable carry-on weight by measuring its micro-Doppler signatures during hovering in an anechoic chamber, benefiting from high SNRs. A continuous-wave (CW) radar operating at a tunable carrier frequency is used, allowing us to explore optimal conditions for target detection. While regulatory considerations are beyond the scope of this study, the approach offers insights into the development of task-specific detection systems. In the next phase, the drone is placed in a wind tunnel, where both payload and wind conditions can be precisely controlled. The CW radar is positioned inside the tunnel to capture the micro-Doppler response under these varying conditions. The results demonstrate deterministic and distinguishable spectral patterns associated with different payloads and wind scenarios, paving the way for the development of robust and comprehensive classification frameworks.

II. ANALYSIS OF ANGLE- AND FREQUENCY-DEPENDENT SCATTERING PROPERTIES OF A DRONE BLADE

The following analysis is general and can be applied to a wide range of configurations. Figure 1(a) and (b) provide a close-up view of a rotor from the DJI Mini 2 drone. Although this model features four independent propellers, we focus on a single rotor to illustrate the concept. While the motor itself contains rotating components, these are electromagnetically shielded and do not contribute significantly to the micro-Doppler signature.

The first experimental assessment focuses on micro-Doppler generation by the blades, considering two angular orientations, namely parallel (0°) and perpendicular (90°) to the vector of propagation k of the incident wave. The $S_{11}$ parameter spectrum (complex reflection coefficient) was measured in the 2–12 GHz frequency range for two orientations, as appears in Figures 1(c) and (d). The measurements were conducted in an anechoic chamber using a broadband horn antenna (IDPH-2018S/N-0807202, 2–18 GHz) and an N5232B PNA-L Microwave Network Analyzer (300 kHz–20 GHz). The absolute values of the Radar Cross Section (RCS) were extracted using a calibration target, specifically a brass disk in this case. The RCS spectra for the two orientations are shown in Figure 1(e). The differential RCS (the difference between the RCS at 0° and 90°) changes, depending on the frequency, from 0.1-0.3 $m^2$ at C-band to 1.5 $m^2$ at X-band. Based on publicly available specifications from press releases [55], [56], [57], this suggests that drone classification can be performed at distances exceeding 1 km under open-sky conditions.

A full-wave numerical analysis was performed using CST Microwave Studio to obtain a comprehensive view of the scattering spectra. An STL 3D model of a plastic propeller was adopted [58], resembling, to some extent, the blade used in our experiment. For conceptual purposes, the exact details of the non-resonant object are not crucial. The structure exhibits nontrivial geometry, dictated by aerodynamic requirements. The main dimensions are 55 mm in length, 9 mm in width, and approximately 1 mm in average thickness. The material parameters of the rotor were taken from [58], [59], with a permittivity of $\varepsilon = 2.8$ and a dielectric loss tangent $\tan(\delta) = 0.0054$.

Figure 1(f) presents the color map demonstrating the numerically calculated electric field intensity (depicted in color) with the horizontal axis representing the angle of rotation and the vertical axis showing frequency, with a broader spectral range considered. Several interesting observations can be made. First, the strongest response occurs around 20 GHz. Since the 24–24.25 GHz band is an unlicensed frequency range, performing micro-Doppler classification at this frequency is appealing, though this claim has only been verified for the specific blades discussed here. However, it is important to note that primary detection is based on its main Doppler shift, which is associated with the center of mass motion. Therefore, the detection and classification problems should be considered together. For example, the DJI Mini 2 drone exhibits maximum RCS at ~3 GHz [55]. The blades exhibit a peak at 20 GHz, which is associated with the internal resonance of the structure, acting as a dielectric resonator with dimensions comparable to the wavelength. Another observation from the colormap is that the maximum scattering is not necessarily obtained when the blade is aligned with the field polarization. At certain frequencies, an additional resonance appears, further complicating the angular and frequency dependence of the scattering. To showcase this behaviour, several angular-dependent RCS values are presented in Figures 1(j-g), obtained by extracting the color map at representative frequencies, as indicated in the caption. To recap, as the angle is a direct function of time (the link is the angular velocity of the blade), the angular dependent modulation of the signal after the Fourier transform becomes a micro-Doppler frequency at the baseband. It is worth noting, however, that capturing these aspects requires very high resolution, which can only be achieved with a Doppler radar and extended time-on-target. In practical scenarios, the resolution is much lower, and only the main Fourier components are detectable, as will be evident in the next section.

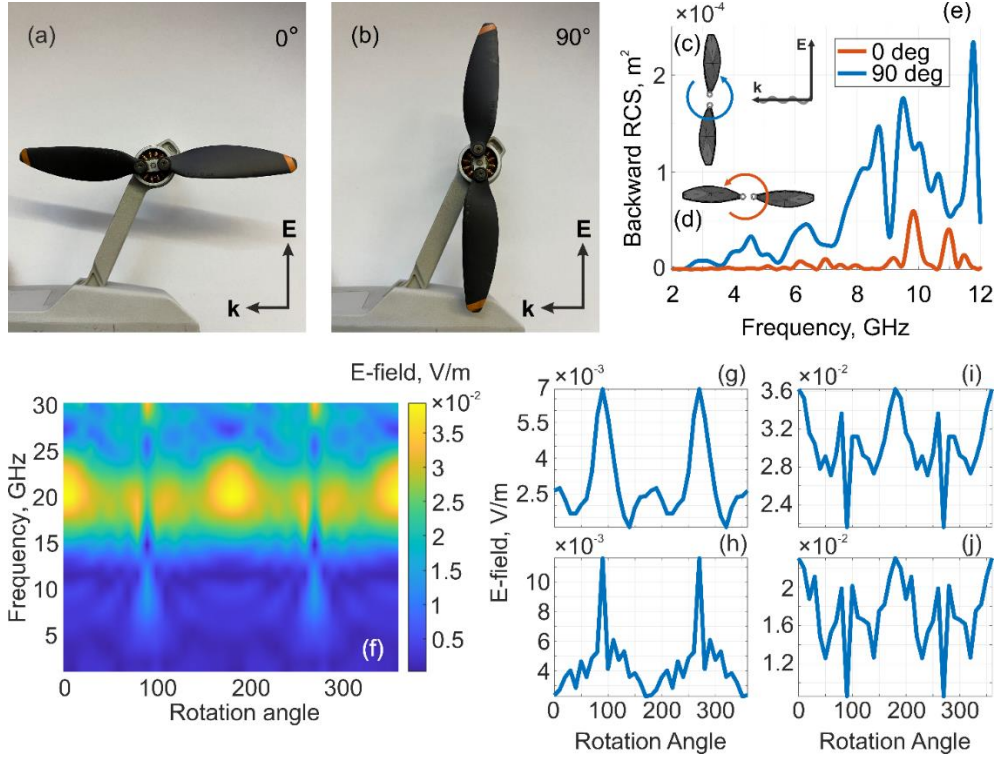

**Fig. 1.** Micro-Doppler analysis of a single blade. (a, b) Photographs of the DJI Mini 2 rotor and blades at two different orientations. (c, d) Interaction scenarios with blades, rotated at 0° and 90° to the vector of propagation k. Cartoon arrows demonstrate the rotation direction. (e) Experimentally measured backward RCS for these two orientations, as indicated in panels c and d. (f) Numerical modeling of an electric field intensity, shown in a color map as a function of frequency and rotation angle. (g–j) electric field intensity as a function of the rotation angle at the frequencies 5 GHz (g), 12 GHz (h), 18.5 GHz (i), and 25 GHz (j), respectively.

## III. Drone in the Anechoic Chamber

The first set of experimental studies was conducted in a controlled indoor environment to disentangle the contribution of the carry-on weight from weather conditions and environmental noises, which reduce signal-to-noise ratio conditions. The photograph of the experimental arrangement appears in Figure 2(a) - a hovering drone (DJI Mini 2) is placed in front of an antenna. To observe the weight-dependent behaviour, a foam box filled with sand was mounted on the top of the drone (Figure 2(b)). This arrangement allows the carry-on weight to be controlled almost continuously. The measurement setup uses a continuous-wave (CW) radar system, which includes Keysight's Performance Network Analyzer P9374A. The co-located transmit and receive antennas are broadband IDPH-2018S/N-0807202 horns. The maximum output from the PNA (20 dBm) is used for the transmit signal. A low-noise amplifier with 30 dB gain is externally applied to the receive chain. The system is controlled via MATLAB, and the time trace of the complex-valued $S_{11}$ parameters is recorded. A typical post-processed signal in the frequency domain (baseband, after downconverting from the 3 GHz carrier) is shown in Figure 1(c). A micro-Doppler peak at approximately 340 Hz at 3 GHz carrier is observed, which corresponds to the doubled frequency of the rotor's rotation (using an unloaded drone for this demonstration). It is worth noting that, instead of a micro-Doppler comb with equidistant peaks (representing odd and even harmonics of the rotor's rotational frequency [56], [57]), only one main contribution is observed. As discussed above, this is due to the SNR in detection, which, even in the case of the anechoic chamber, is insufficient to resolve secondary features.

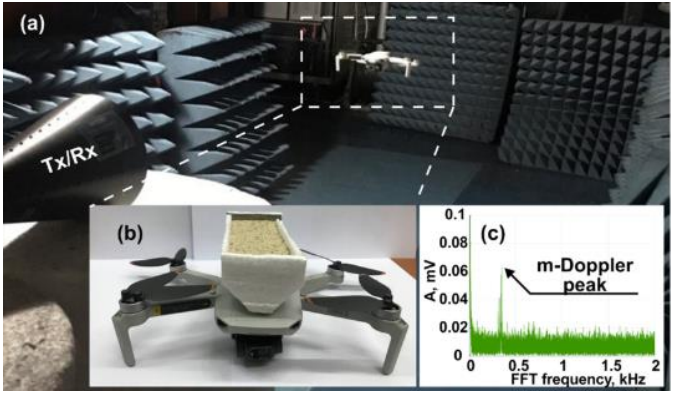

Fig. 2. Experimental setup for retrieving the drone's weight from micro-Doppler signatures. (a) Photograph of a drone (DJI Mini 2), hovering in an anechoic chamber. (b) Photograph of a drone with a carry-on weight – a foam box, filled with sand. (c) Baseband response of a CW (3GHz carrier) radar, indicating the presence of a micro-Doppler peak. The drone was unloaded.

Several representative frequencies, i.e., 2, 3, 4, 7, 8, and 10 GHz, were chosen for consideration. The analysis involves collecting the backscattered signal from the drone (10 seconds on target), down-converting it to the baseband, and performing an FFT analysis. Figure 3 presents color maps corresponding to different carrier frequencies. Vertical slices represent baseband spectra for each carry-on weight (x-axis). It is important to note that these plots should not be confused with spectrograms obtained from swept FFT used in time-domain analyses. In this case, the drone is hovering, and the process is in steady state. When comparing different carriers, the most pronounced response is observed at 7 GHz, which is consistent with the results shown in Figure 1. In all cases, only a single micro-Doppler frequency is detected, and other harmonics in the comb are not visible, similar to the results discussed in Figure 2 (c). Focusing on the 7 GHz case (Figure 3(d)), the weight-dependent behavior of the peak position becomes evident. As the carry-on weight ($w$) increases from 0 to 120 g (the maximum weight the specific drone can safely lift), the micro-Doppler frequency shifts continuously from 340 to 430 Hz. This shift is approximately linear, following the empirical equation:

$$f_{MD}(w) = 340[Hz] + 90[Hz]\frac{w[g]}{120[g]}. \quad (1)$$

To reiterate, the micro-Doppler spectra do not directly depend on the carrier frequency, as the effect is described by amplitude modulation of the radar echo signal (also evident from different panels in Fig. 3). However, the SNR does depend on the carrier frequency, as scattering efficiency is influenced by factors such as the size, material, and positioning of the rotors. An additional observation is that the micro-Doppler response of this particular drone is barely detectable at X-band, which is of primary importance to certain classes of surveillance radars. Note that while the results in Figure 1 are obtained with a single blade, the results in Figure 3 are for the entire drone, which explains the observed shift in frequency. It is also worth noting that for this specific drone, the 24 GHz band is a preferable choice, as discussed above.

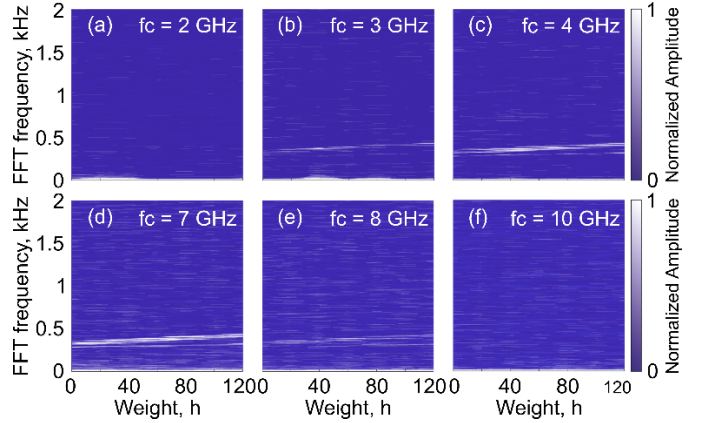

Fig. 3. Monitoring the drone's carry-on weight using micro-Doppler frequency shift. Vertical slices through the color maps represent baseband spectra, with carry-on weights shown along the horizontal x-axis. The carrier frequencies of the CW radar are indicated in the plots.

IV. DRONE IN THE WIND TUNNEL

While the impact of weight on the micro-Doppler signature is evident from the previous section, similar effects can arise if the drone is influenced by wind. It is important to note that radar analysis enables the extraction of the drone's flight velocity, offering insight into how more complex flight scenarios, affected by additional factors, can be resolved in the process of extracting the payload. To recap, the effect of drone flight velocity is not considered in this study due to the practical limitations of the measurement setup, and the focus is on the hovering drone. As will become evident, weight and wind have different impacts on the micro-Doppler signatures of the hovering drone. As a result, these two effects can be disentangled, providing a method for extracting carry-on weight information under any flight conditions without ambiguity.

Figures 4(a) and (b) demonstrate the photos from the facility used in this experiment. We used an open-loop wind tunnel with a cross-section of 1.5 m x 1 m (width x height) and an optically accessible test section of 2 m, located downstream of the radar at 7.5 m. This open-loop wind tunnel uses a DC motor to run a blowing-down fan, creating wind speeds of up to 6 m/s. The bottom wall is installed with additional canopy-like roughness to create a thicker boundary layer mimicking an atmospheric boundary layer with larger and stronger turbulent coherent structures above the roughness.

In these experiments, the payload weights were 3D printed from PLA plastic, forming lego-like parallelepipeds with varying weights, and were securely attached to the drone body (see Figure 4(c)). The radar is deployed directly inside the tube with the same hardware and processing routine as described in the previous section.

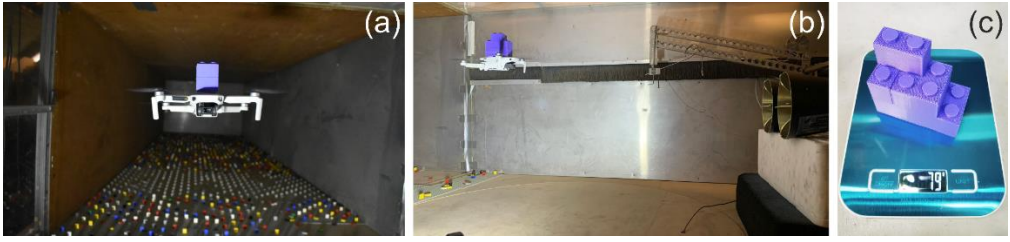

**Fig. 4.** Drone measurements in the wind tunnel. (a) Photograph of the drone hovering in the wind tunnel. The wind is always coming from one side of the drone; in this photo, it's coming from the back. The colored cubes at the bottom are used to create a stronger boundary layer from the bottom, mimicking, to some extent, the stronger turbulence in the atmospheric boundary layer environment below the drone. (b) The drone is positioned against the Tx and Rx radar antennas. The wind is from left (upstream) to right (downstream). (c) Calibration of payload weight.

The main result, presented in the form of a color map (not a spectrogram), is shown in Figure 5. In the first set of experiments, a rear wind has been applied. The vertical axis on this plot represents the micro-Doppler frequency. The plot consists of 7 sections, each corresponding to a different carry-on weight, as indicated in the insets. The horizontal axis of each subplot corresponds to the wind velocity, ranging from 0 to 3.3 m/s. The experiments were conducted with a CW radar operating at 7 GHz. Each wind-weight condition on the plot was assessed by observing the target for 10 seconds. Several observations can be made. First, the main micro-Doppler signature splits into two branches, with the separation distance increasing as the wind velocity rises. In the case of zero wind velocity, there is almost no splitting. The upper branch of the plot (using only points corresponding to zero wind) follows Equation 1 quite accurately. Second, the observed splitting follows the front wind velocity linearly, allowing for the extraction of an empirical relation as follows:

$$f_{MD}^{UP}(w, v) = f_0 + 4v[m/s] + 0.75w[g],$$
$$f_{MD}^{DOWN}(w, v) = f_0 - 12v[m/s] + 0.75w[g], \quad (2)$$

where $f_0$ is the micro-Doppler frequency as generated by an unloaded hovering drone under zero-wind conditions, $w$ is the payload in grams, and $v$ is the front wind in m/s.

The most remarkable result is the frequency difference between the upper and lower branches, which does not depend on the payload at all, and follows:

$$\Delta f = f_{MD}^{UP}(w, v) - f_{MD}^{DOWN}(w, v) = 16v[m/s]. \quad (3)$$

Equation 3 shows that the wind-weight ambiguity can be removed. Knowing the wind speed allows it to be used as a parameter in Equation 2, enabling the extraction of the weights. To recap, from the up-shifted micro-Doppler frequency $f_{MD}^{UP}$, in Equation 2, the drone's total payload can be directly estimated. However, ambient wind produces an almost identical, linear shift in $f_{MD}^{UP}$, leading to an ambiguity between wind and weight effects. For example, in Figure 5, a headwind of 3.3 m/s induces the same frequency shift as carrying an extra 15–20 g under zero-wind conditions. To resolve this, Equation 3 models the frequency difference, $\Delta f$, explicitly as a function of wind speed. By first using Equation 3 to extract the wind speed from the observed $\Delta f$, one can then apply Equation 2 to the wind-corrected frequency and obtain an unambiguous estimate of the onboard mass.

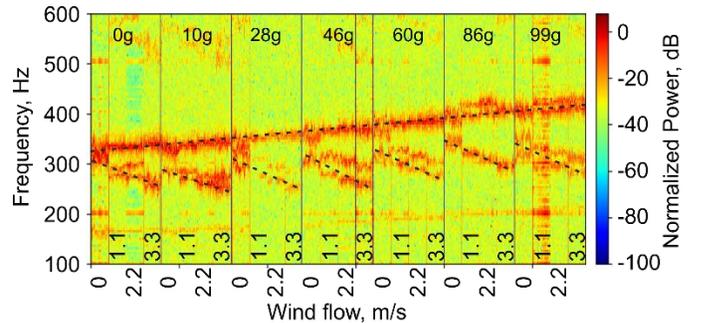

**Fig. 5.** Wind velocity-payload relation color map. The 7 sections of the plot represent different carry-on weights, as indicated in the insets. The vertical axis represents the micro-Doppler frequency in Hz, while the horizontal axis corresponds to the front wind velocity, ranging from 0 to 3.3m/s, for each of the 7 subplots. Micro-Doppler branching is highlighted by black trendlines.

The physics behind the splitting is related to the flight controller, and the way the drone counteracts weight and wind differs. When the payload is balanced, it imposes an equal load on all four rotors. However, the impact of the wind is different. The wind attempts to displace the drone, and to counteract this, the drone tilts. As a result, the forward and rear rotors (regarding the wind direction) begin rotating at different velocities to maintain the tilt angle of the drone's body to the wind direction and fixed position. Consequently, two prominent micro-Doppler peaks appear from the pair of blades, as seen in Figure 5.

The data from Figure 5 can be presented differently, offering a more detailed and comparative insight into the results. Figure 6(a) presents the baseband spectra (using a 7 GHz carrier) for a constant front wind (ISO-wind) and varying carry-on weights. Figure 6(b) shows the variation of the spectra with a constant weight (ISO-weight) while the wind changes, as another way to present the results. From this type of analysis, the branching of the Doppler frequencies is clearly observed. Additionally, the analysis reveals that the required Doppler resolution should be on the order of 10 Hz. A 10 Hz Doppler resolution can be achieved with radars highlighted in works [60]. The peaks also become broader under certain conditions, which can be attributed to instabilities as the flight controller attempts to mitigate non-optimal conditions.

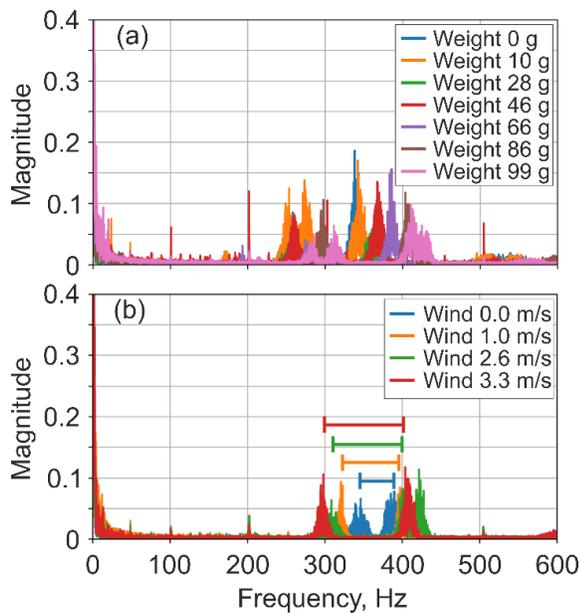

**Fig.6.** (a) ISO-wind micro-Doppler signatures. The weight parameters are indicated in the legends. The wind is 3.3 m/s. (b) ISO-weight micro-Doppler signatures. The wind parameters are indicated in the legends. The payload is 66g.

V. OUTLOOK AND CONCLUSION

In this study, we demonstrated that micro-Doppler analysis provides a reliable method for disentangling the effects of wind and payload weight on a hovering drone's dynamics. By analyzing the micro-Doppler signatures of rotating blades, we identified distinct spectral patterns associated with changes in payload and wind conditions. The systematic experiments conducted in both an anechoic chamber and a wind tunnel showed that the effects of weight and wind could be quantified separately, providing a pathway for accurate monitoring of drone payloads in real-time. The proposed deterministic algorithm, based on the branching of the micro-Doppler spectra, offers a robust solution for extracting payload information, even in the presence of varying wind conditions. This work lays the foundation for future radar-based drone monitoring systems that could play a crucial role in various applications, including security, logistics, and air traffic management. Moreover, integrating deterministic physics-based models with machine learning techniques promises to enhance the reliability and interpretability of future radar detection algorithms.